\def\lesssim{\mathrel{\hbox{\rlap{\hbox{\lower3pt\hbox{${\sim}$}}}\hbox{\raise2pt\hbox{$<$}}}}}

\documentclass[useAMS,usenatbib]{mn2e}
\usepackage{cleveref}
\usepackage{natbib}
\usepackage{graphicx}
\usepackage{amsmath}
\usepackage{epstopdf}

\newcommand{\msun}{M$_\odot$}
\newcommand{\rsun}{R$_\odot$}

\newcommand{\ms}{$M_\odot$}

\def\lesssim{\mathrel{\hbox{\rlap{\hbox{\lower3pt\hbox{$\sim$}}}\hbox{\raise2pt\hbox{$<$}}}}}
\def\gtrsim{\mathrel{\hbox{\rlap{\hbox{\lower3pt\hbox{$\sim$}}}\hbox{\raise2pt\hbox{$>$}}}}}
\def\gtreq{\mathrel{\hbox{\rlap{\hbox{\lower3pt\hbox{$-$}}}\hbox{\raise2pt\hbox{$>$}}}}}

\title[Fall-Back Discs in Common Envelope Binary Interactions]{Considerations on the Role of Fall-Back Discs in the Final Stages of the Common Envelope Binary Interaction}
\author[Rajika L. Kuruwita, Jan Staff \& Orsola De Marco]{Rajika L. Kuruwita$^{1,2,3}$, Jan Staff$^{1,2,4}$, Orsola De Marco$^{1,2}$\\ $^{1}$Department of Physics and Astronomy, Macquarie University, Sydney, NSW, 2109,  Australia\\ $^{2}$Astronomy, Astrophysics and Astrophotonics Research Centre, Macquarie University, Sydney, NSW, 2109, Australia\\ $^{3}$Current address: Research School of Astronomy and Astrophysics, Australian National University, Canberra, ACT 2611, Australia\\ $^{4}$Current address: Department of Astronomy, University of Florida, Gainesville, FL 32611, USA}

\begin{document}

\date{}

\pagerange{\pageref{firstpage}--\pageref{lastpage}} \pubyear{2002}

\maketitle

\label{firstpage}

\begin{abstract}
The common envelope interaction is thought to be the gateway to all evolved compact binaries and mergers. Hydrodynamic simulations of the common envelope interaction between giant stars and their companions are restricted to the dynamical, fast, in-spiral phase. They find that the giant envelope is lifted during this phase, but remains mostly bound to the system. At the same time, the orbital separation is greatly reduced, but in most simulations it levels off at values larger than measured from observations. We conjectured that during the post-in-spiral phase the bound envelope gas will return to the system. Using hydrodynamic simulations, we generate initial conditions for our simulation that result in a fall-back disk with total mass and angular momentum in line with quantities from the simulations of Passy et al. We find that the simulated fall-back event reduces the orbital separation efficiently, but fails to unbind the gas before the separation levels off once again. We also find that more massive fall-back disks reduce the orbital separation more efficiently, but the efficiency of unbinding remains invariably very low. From these results we deduce that unless a further energy source contributes to unbinding the envelope (such as was recently tested by Nandez et al.), all common envelope interactions would result in mergers. On the other hand, additional energy sources are unlikely to help, on their own, to reduce the orbital separation. We conclude by discussing our dynamical fall-back event in the context of a thermally-regulated post-common envelope phase.
\end{abstract}

\begin{keywords}
Stellar Evolution -- Binary stars: Common Envelope.
\end{keywords}

\section{Introduction}
\label{sec:introduction}

The common envelope binary interaction was proposed by \citet{paczynski_common_1976} to explain the existence of cataclysmic variables and evolved, close binary star systems. A common envelope event is supposed to occur when a binary system with a sufficiently small mass ratio ($M_{\rm accretor}/M_{\rm donor}$) comes into contact, usually as the result of one star expanding as it evolves into a giant. If mass transferred from the expanding star cannot be accommodated by the companion, then the companion would quickly enter a common envelope, in-spiral towards the core of the primary and cause the removal of the envelope via the transfer of orbital energy and angular momentum to the gas. The result would be a close binary composed of the core of the giant and the companion, or, if the envelope is not removed, a merger (for a review see \citealt{ivanova_common_2013}).

The common envelope interaction has been investigated via 3-dimensional hydrodynamic simulations using both smooth particle hydrodynamics (SPH) and eulerian grid codes, and all have found that by the time the fast in-spiral slows down, the envelope is lifted but not completely unbound. \citet{rasio_formation_1996}, using an SPH code, found that approximately 90 per cent of the envelope of their 4~\msun\ red giant branch (RGB) star remains bound. \citet{sandquist_double_1998}, using a stationary nested grid method with intermediate mass giants found that approximately 60 per cent of the envelope was still bound to the system. \citet{ricker_interaction_2008} and \citet{ricker_amr_2012} improved on this simulation by using an adaptive mesh refinement grid code. With this method they found that 75 per cent of the envelope of a 1-\msun, early RGB star remained bound to the system. \citet{passy_simulating_2012}, comparing single grid and SPH simulations concluded that approximately 85 per cent of the envelope of 1-\msun, late RGB star remains bound to the system. In all these simulations most of the envelope is lifted away from the orbiting cores, to  approximately 1000~\rsun\ in the case of the SPH simulations of Passy et al. (2012), but remains mostly bound to them.

A number of other mechanisms have been proposed to assist in the removal of the bound gas. Passy et al. (2012) estimated that radiation pressure alone would be unlikely to remove the remaining bound mass. A more promising mechanism, is that expanding gas cools and recombines and that recombination energy may contribute to unbinding the bound gas \citep{han_formation_1995,ivanova_role_2015}. \citet{nandez_recombination_2015} included the recombination energy via the equation of state. They simulated a common envelope interaction between compact (R$\sim$16-30~\rsun), low mass (M=1-1.8~\ms) RGB primaries and a 0.36~\ms, compact companion with an SPH code. These simulations completely removed the envelope, while using an ideal gas equation of state removed only 50 per cent of the envelope.
On the other hand, \citet{Ohlmann2016} simulated a more extended giant with the moving-mesh code {\sc arepo} \citep{Springel2010} and preliminary results (Ohlmann, private communication) show that adding recombination energy to those simulations doubles the unbound mass, but the total unbound mass is still only less than a quarter of the total.

	
Some hydrodynamic simulations also tend to produce final orbital separations larger by a factor of a few than observed in post-common envelope binaries \citep{passy_simulating_2012}\footnote{Observed post common envelope binaries tend to have separations of $\lesssim 4$~R$_\odot$ \citep{schreiber_age_2003,zorotovic_post-common-envelope_2010,de_marco__2011}, apparently independently of the companion to primary mass ratio. These small orbital separations are not an effect of orbital decay at the hand of physical mechanisms that take place after the common envelope. The observed systems are either central stars of planetary nebula, which have only just left the asymptotic giant branch, or young white dwarves, for which not enough time has elapsed for further orbital shrinkage.}. More importantly, simulated systems with larger companion to primary mass ratios result in systematically larger separations, a trend also not observed. This said, simulations with heavier and/or more compact envelopes do reproduce observed separations \citep{rasio_formation_1996,Nandez2015}. The short orbital separation achieved by Nandez et al. (2015) is likely disconnected from the implementation of the recombination energy formalism and is due instead to the compact nature of the primary. The SPH simulations of \citet{rasio_formation_1996}, that did not include the effects of recombination energy, also achieved a very small orbital separation (1~\rsun), possibly because of their relatively compact (62~\rsun) and more massive (4~\ms) RGB primary.

It has also been suggested that the bound gas from the interaction returns to the binary, and interacts with the core of the giant and the companion again. \citet{Kashi2011} calculated analytical scenarios where 1-10 per cent of a common envelope would return, making, in their case, a 0.2~\msun\ stable disc around the binary, which would then interact tidally with it, thereby shortening the orbital separation. Such disks would not interact directly with the binary. Examples of circumbinary disks around post-common envelope binaries include those detected around post-RGB and post-AGB star binaries by, e.g., \citet{vanWinckel2009}, or those around suspected central stars of planetary nebula binary mergers \citep{DeMarco2002,DeMarco2002b}.
	
In this paper we consider instead what might happen if a substantial fraction of the envelope remained bound and fell back onto the binary in the form of a disk or torus. Although some of the gas would possess sufficient angular momentum to form a circumbinary disk, some would likely crash onto the central binary, as was estimated analytically by \citet{tocknell_constraints_2014}. A new interaction phase may provide the opportunity to further reduce the orbital separation, thereby transferring more energy and angular momentum to the gas, which could, as a result, become unbound. 

Here we simulate such a fall-back event, using a grid-based hydrodynamics approach, guided by the binary separation, the bound mass and angular momentum values at the end of the simulations of \citet{passy_simulating_2012}. Our a setup is completely artificial at time zero. It is purely designed to result in an in-falling disk with the correct parameters shortly after the start of the simulation. We use these simulations as the basis of a discussion of the possible role played by gas falling back onto the binary.

In Section~\ref{sec:theinitialsetup} we describe the simulation setup. In Section~\ref{sec:results} we discuss our results: the orbital separation decrease achieved (Section~\ref{ssec:theorbitalseparation}), the amount of unbound gas (Section~\ref{ssec:determinationoftheamountofunboundmaterial}), the conservation of energy and angular momentum in our simulations (Section~\ref{ssec:energyandangularmomentumconservation}), the impact of the gas temperature on the outcome (Sections~\ref{ssec:impactofthegastemperatureontheparticlesinspiral}) and the effects of numerical resolution (Section~\ref{ssec:numericalresolution}). In Section~\ref{sec:discussion}, guided by our simulations, we discuss the likely effects of fall back discs on orbital separation and gas unbinding. We summarise and conclude in Section~\ref{ssec:furtherconsiderations}.

\section{The Simulation Setup}
\label{sec:theinitialsetup}

The simulations are carried out with the grid-based hydrodynamic code \emph{Enzo}, adapted to simulate the common envelope interaction by \cite{passy_simulating_2012}. This code includes gravity and hydrodynamics calculated using a ZEUS solver. We use an ideal gas equation of state ($\gamma = 5/3$) in a cartesian grid with $256^3$ cells, corresponding to $143$~R$_\odot$ in each direction. A smoothing length of 1.5 cells is implemented \citep{passy_simulating_2012}. In the simulations carried out by \citet{passy_simulating_2012}, a $1$~M$_\odot$ main sequence star was evolved to the RGB using the stellar evolution code {\sc EVOL} \citep{herwig_evolution_2000}. This produced an RGB star with a radius of $83$~R$_\odot$, a total mass of $0.88$~M$_\odot$, and a core mass of $0.39$~M$_\odot$. The common envelope interaction with a $0.6$~M$_\odot$ companion resulted in a final separation of  $20$~R$_\odot$. Approximately $0.44$~M$_\odot$ of the RGB star's envelope remained bound to the system. These final conditions are used here to construct our simulation of a fall back event. 

Our simulations begin with two point mass particles of mass $0.39$~M$_\odot$ and $0.6$~M$_\odot$ representing the core of the RGB star and the companion, with an initial separation of $20$~R$_\odot$. { Ideally we would like to follow the simulation of Passy et al. (2012) further in time, and wait for the fall back event to happen naturally. However, this would be computationally expensive and would require a considerably larger computational domain. While this is not strictly beyond the realm of possibility, we start with a much simpler model setup from which future simulation can gain insight.

What we know from Passy et al. (2012) is that 0.44~\msun\ of gas will fall back onto the central binary and that the total angular momentum of the bound material is $1.5 \times 10^{52}$~g~cm$^2$~s$^{-1}$ and is directed in the $z$-direction, perpendicular to the orbital plane.  

To encourage the gas to fall onto the central system as a disk and to ensure that the in-falling gas has the appropriate amount of mass and angular momentum, we start with a constant density of $10^{-6}$~g~cm$^{-3}$. This value is chosen because in this way the total gas mass on the grid at the beginning of the simulation (0.49~\msun) is similar to the mass of the bound material in the simulation of \citet[][0.44~\msun]{passy_simulating_2012}, which we expect to fall back. A bonus is that this is also similar to the density in which the core and companion are immersed at the end of the simulation by Passy et al. (2012; $0.5 \times 10^{-6}$~g~cm$^{-3}$).

The gas on the equatorial plane ($z=0$) is given an orbital velocity around the binary corresponding to 0.75 times the Keplerian value. Gas situated above and below the equatorial plane ($z$ values larger or smaller than zero) was given the same velocity as the gas with the same $x$ and $y$ coordinates. This setup is equivalent to a set of concentric, solidly rotating cylinders. In this way the gas on or close to the orbital plane, which has sub-Keplerian velocities falls onto the particles, while gas farther away from the plane has super-Keplerian velocities and is evacuated. This is an expedient to generate a fall-back disk. Already by 0.02 years into the simulation, the gas has redistributed itself into a disk. 
\begin{figure}
\centering
\includegraphics[width=0.4\textwidth]{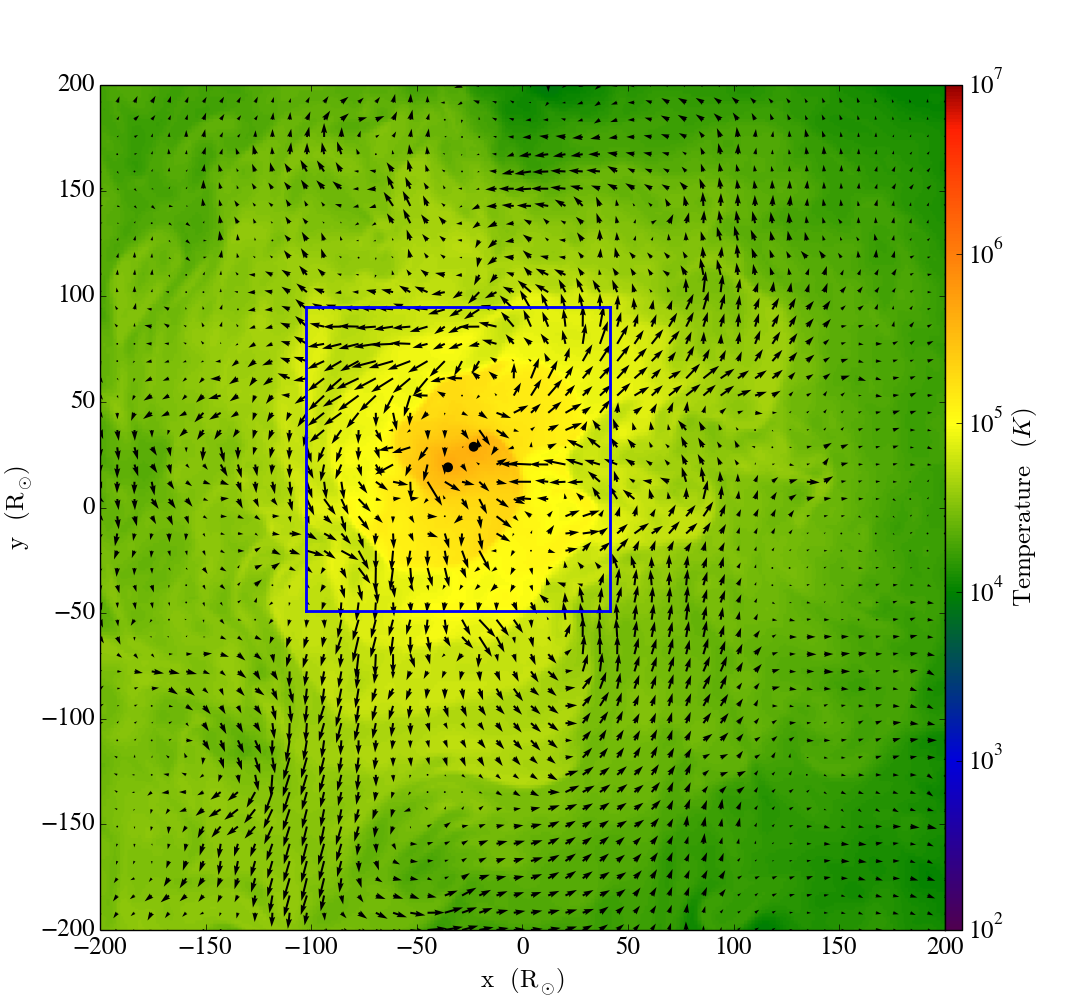}
\caption{{ A temperature slice along the orbital plane, from the simulation of \citet{passy_simulating_2012} for the Enzo7 simulation with a $0.6$~M$_\odot$ companion. This slice is from the end of the simulation after 1000 days. The box indicates the size of the computational domain used in the simulations of this paper.}}
\label{passy_temperature}
\end{figure}

One issue that cannot be resolved presently is that of the temperature of the fall-back disk. Since this issue does play a role on the type of disk that forms, or even if a disk would form at all, we dwell on it below.

\subsection{The disc's temperature}
\label{ssec:thediscsmassandangularmomentum}


The temperature of the fall back disk is unknown. The temperature at the end of the simulation of Passy et al. (2012) is shown in Fig.~\ref{passy_temperature}. The volume weighted average temperature in the entire simulation box is 25\,000~K, while in the smaller box with the size of our simulation domain it is $\sim$100\,000~K. The mass-weighted average over our small simulation domain is $\sim$120\,000~K, similar to the volume-weighted average, due to a relatively homogeneous mass density over the small central volume. Close to the particles the temperature is closer to 1 million degrees Kelvin, and possibly it would be higher if nuclear energy generation had been allowed to contribute.  By the end of their simulation, the computational volume only samples low amounts of left over, out-streaming gas (0.04~\msun). Most of their envelope mass is, at that time, outside the simulation domain. 

The end of the simulation of Passy et al. (2012) cannot therefore be used directly to constrain the temperature of in-falling gas. Although we have conjectured that the gas will fall back organised in a disc, we admit to the fact that, depending on the temperature behaviour of the in-flowing gas the structure that forms may become pressure supported before it becomes rotationally supported. If this were so an actual disk may not form in the way we have envisaged. Here we nonetheless experiment with the idea that a disk does indeed form, but we admit to this caveat and leave this question ultimately unanswered, suggesting that it requires a full simulation of the post-dynamic in-fall phase, such as those attempted by \citet{Reichardt2016}.

The gas would adiabatically cool as it expands and heat as it falls back. Some radiative cooling could also take place during the expanded phases, resulting in lower temperatures of the fall-back material. It is likely that the temperatures would be quite high, close to what they were in the inner regions of the star. 

We also should point out that the temperature which we give the gas at $t=0$ is not the temperature of the fall-back disk. Pre-empting our results (Section~\ref{sec:results}), the value of the temperature changes very quickly as the disk falls into place and the interaction between the in-falling gas and the particles injects energy. Ultimately no matter what the temperature at the start of the simulation, it converges towards  a volme-averaged virial value of $\sim$50\,000~K at about 0.3~yr after the start of the simulation \citep{Kuruwita2015}. What does change for different initial values of the temperature is, on the other hand, the mass of the forming disk: the higher the temperature the lower the mass is for the same velocity profile. This may argue that a higher temperature value would result in more pressure support and a fall-back event that looks less like a disk and more like an envelope.

Under the assumption that a disk-like structure does form, and that the mass and angular momentum of this structure should be as those dictated by the simulation of Passy et al. (2012), we carry out three simulations, with two initial temperature values: one \emph{Cool} simulation and two \emph{Hot} simulations. Our \emph{Cool} simulation begins with a uniform temperature profile of $350$~K, while the initial temperature of the \emph{Hot} simulations is 35~000~K. The second {\it Hot} simulation has a slower velocity profile and is designed to maintain the same amount of disk mass as the {\it Cool} simulation but with a higher temperature. We discuss temperature further in Section~\ref{ssec:impactofthegastemperatureontheparticlesinspiral}.

\begin{figure}
\centering
\includegraphics[width=0.44\textwidth]{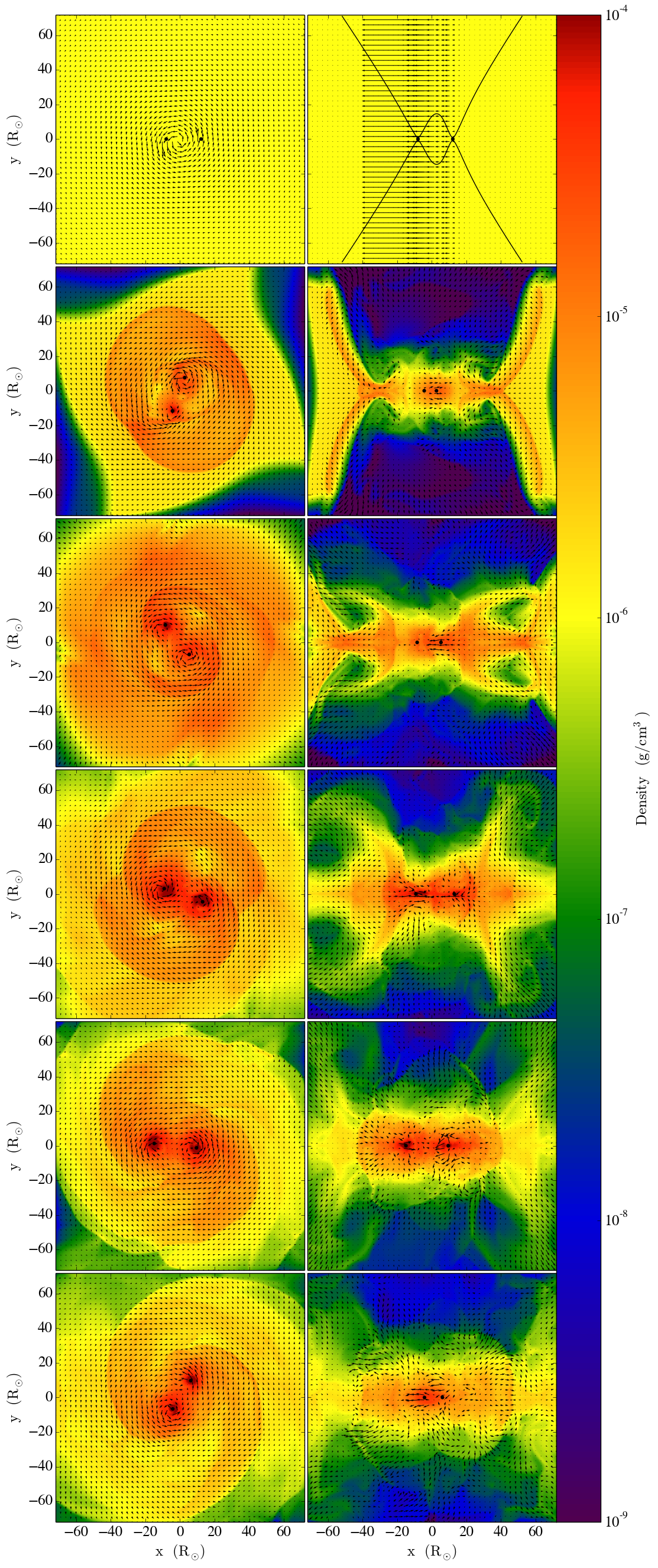}
\caption{Density slices along the orbital plane (left panels) and perpendicular to orbital plane (right panels), at the beginning of the simulation (top row), and then at 0.02, 0.04, 0.06, 0.08 and 0.10 years (following rows). The contour lines in the top-right panel are discussed in the text.}
\label{fallback}
\end{figure}

\begin{figure}
\centering
\includegraphics[width=0.5\textwidth]{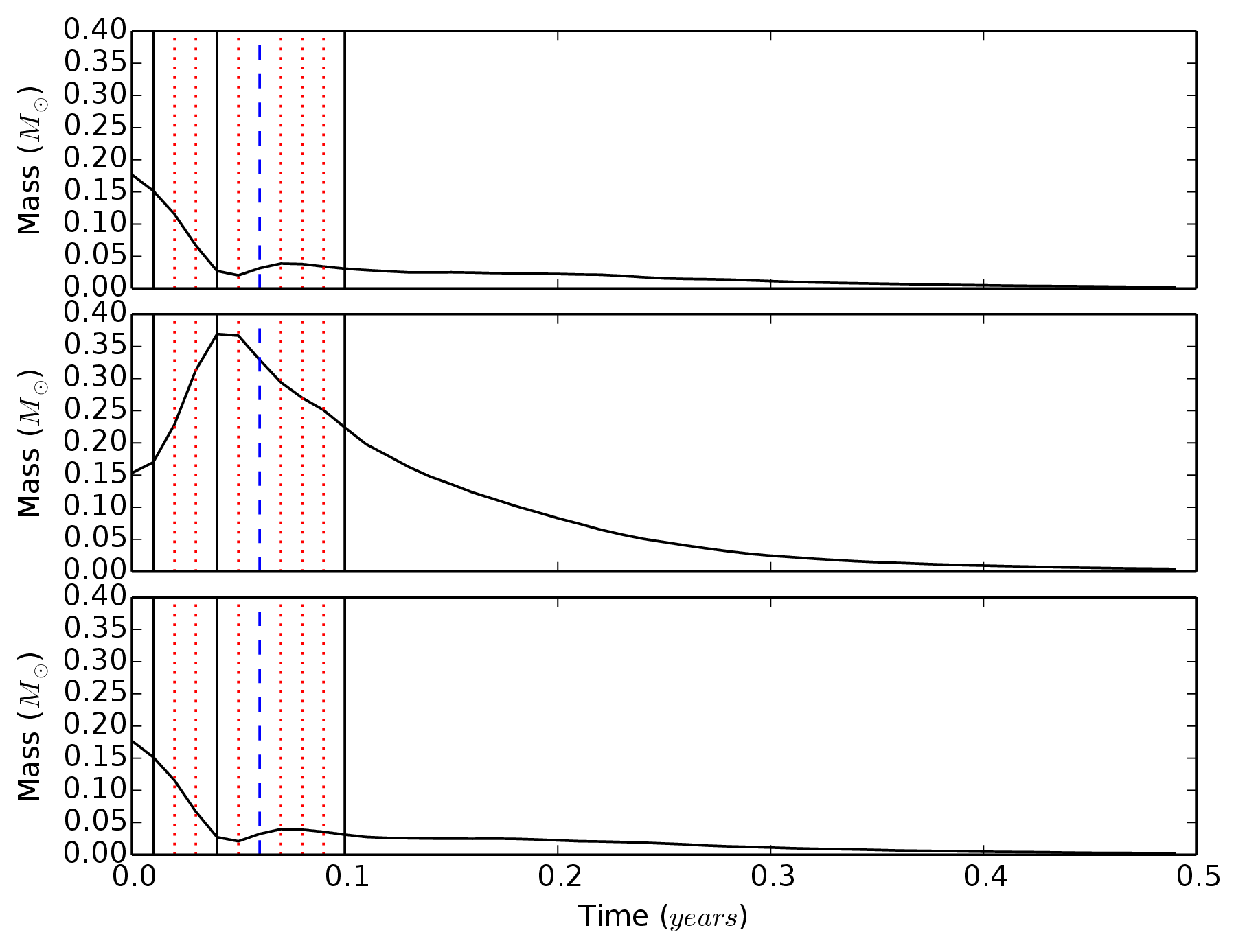}
\caption{Total gas mass within three regions in the grid. The middle panel shows the mass centred on the orbital plane (i.e., $0.35 \le z \le 0.65$, where 1.0 is the dimension of the grid). Top and bottom panels show the mass above and below the central region. The vertical lines are time reference points and are described in Section~\ref{ssec:thediscsmassandangularmomentum}.}
\label{split_mass}
\end{figure}

\begin{figure}
\centering
\includegraphics[width=0.5\textwidth]{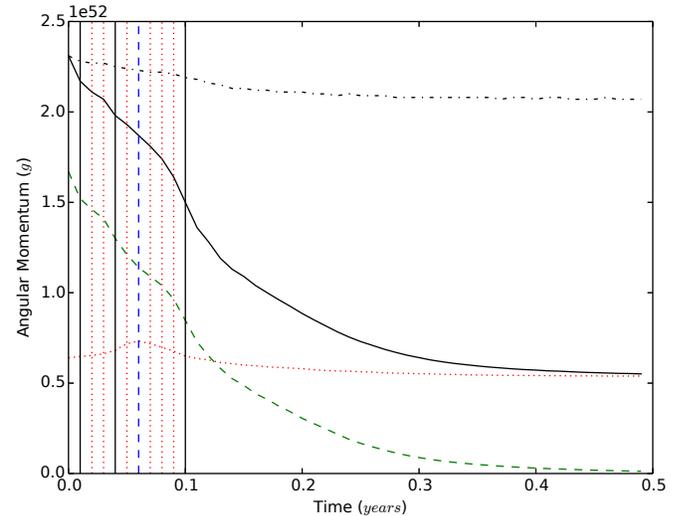}
\caption{The evolution of the angular momentum components throughout the simulation. The solid black line is the total angular momentum of the system. The green dashed line is the angular momentum of the gas and the red dotted line is the angular momentum of the particles. The dot-dash black line is the total angular momentum, corrected for mass losses from the computational domain. The vertical lines are time reference points and are described in Section~\ref{ssec:thediscsmassandangularmomentum}.}
\label{angular_momentum_timeseries}
\end{figure}

\section{Results}
\label{sec:results}

Figure \ref{fallback} shows the first 0.1 years of the fall-back event. In the top right panel of this figure we show a contour that indicates the locus where the gas velocity is equal to the Keplerian value. The Keplerian velocity is calculated using $v_{\rm Kep} = \sqrt{- \phi}$, where $\phi$ is the gravitational potential from the core and companion at a given point in the grid. Above and below the core and companion, within the cones outlined by these contours, the gas is super-Keplerian, while outside the cone it is sub-Keplerian, so the gas within the cones is rapidly evacuated, while the initially sub-Keplerian gas falls onto the orbital plane and towards the particles. The pocket of gas between the particles extending just above and below them is also sub-Keplerian. The arrows indicate the component of the velocity projected onto the plane of the slice. In the top-right panel of Figure \ref{fallback} the velocity arrows should be zero, but they are not due to the fact that the centre of the grid is along cell boundaries and not cell centres, so we are plotting a slice which is half cell in front of the $y=0$ value. By 0.03 years the density along the orbital plane is approximately 4 orders of magnitude higher than above and below the plane, so we can confidently say we have created a disc.
 
 In Figure \ref{fallback} an ``edge effect" is evident at 0.02 years, particularly on the left column, as low density pockets develop at the computational domain boundary as gas should be drawn from outside of the computational domain due to the velocity field imposed, but the outflow boundary conditions do not allow this to happen. This low density, low temperature, low pressure pockets rapidly equalise and do not have a dynamical effect on the simulation.
 
We estimate the mass of the fall-back disc (Figure \ref{split_mass}) by dividing our grid into 3 sections. The middle section is defined by a box that spans 30 per cent of the $z$-axis and is centred on the orbital plane. The regions above and below contain the remaining computational domain. By creating a mass time series for each of the three regions we can estimate that the fall-back disc contains $\sim0.38$~M$_\odot$, as this is the total mass in the central region shortly after the initial evacuation of gas from the grid at 0.04 years. This value is reasonably close to the $0.44$~M$_\odot$ of bound mass found by \citet{passy_simulating_2012}, which is expected to fall back.

Throughout this paper, time series figures, such as Figure \ref{split_mass}, have 10 vertical reference lines. The blue dashed line indicates what we call the ``unbinding event" in Sec.~\ref{ssec:determinationoftheamountofunboundmaterial} and corresponds to a  time of $0.06$~years. The black solid lines indicate 3 major mass loss events over the course of the simulation. These correspond to times $0.01, 0.04$ and $0.1$~years. The red dotted lines each indicate times of $0.02, 0.03, 0.05, 0.07, 0.08$ and $0.09$~years. These reference lines allow easy comparison between figures.

\citet{passy_simulating_2012} calculated that the total $z$-component of the angular momentum of the bound gas and of the particles at the end of their simulation was $1.5 \times 10^{52}$~g~cm$^2$~s$^{-1}$ and $0.6 \times 10^{52}$~g~cm$^2$~s~$^{-1}$, respectively (see fig. 8 and 9 of \cite{passy_simulating_2012}). In our simulations, the total angular momentum of the gas and particles after the mass loss of the initially super-Keplerian gas at $t \sim 0.04$~years is $\sim 1.3 \times 10^{52}$~g~cm$^2$~s~$^{-1}$ and $\sim 0.7 \times 10^{52}$~g~cm$^2$~s~$^{-1}$, respectively (see Figure \ref{angular_momentum_timeseries}), justifying our choice of initial velocities. 

In what follows we describe the {\it Cool} simulations, carried out with the initial temperature value of 350~K. In Sec.~\ref{ssec:impactofthegastemperatureontheparticlesinspiral} we compare this simulation with the two simulations carried out with a hotter initial temperature.

\subsection{The orbital separation}
\label{ssec:theorbitalseparation}

The {\it Cool} simulation resulted in the decrease of the orbital separation by $35-43$ per cent as shown in Figure \ref{separation_timeseries}. This value range was calculated using the initial orbital separation of 20~\rsun, or using the maximum separation observed after the initial widening of the orbit (24.5~\rsun). This initial increase in separation occurs over 0.08 years, or a couple of orbital periods (Figure \ref{cool_eccentricity}). It is in small part due the system developing an eccentricity of about 0.15 over the same time period but,  more importantly, it is due to the fact that the gas speed at the location of the particles is initially faster than the particle's orbital velocity, thus exerting a temporary drag in the  direction of motion (the gas is transferring orbital energy to the particles). We will discuss this effect further in Section~\ref{ssec:impactofthegastemperatureontheparticlesinspiral}. 

\begin{figure}
\centering
\includegraphics[width=0.5\textwidth]{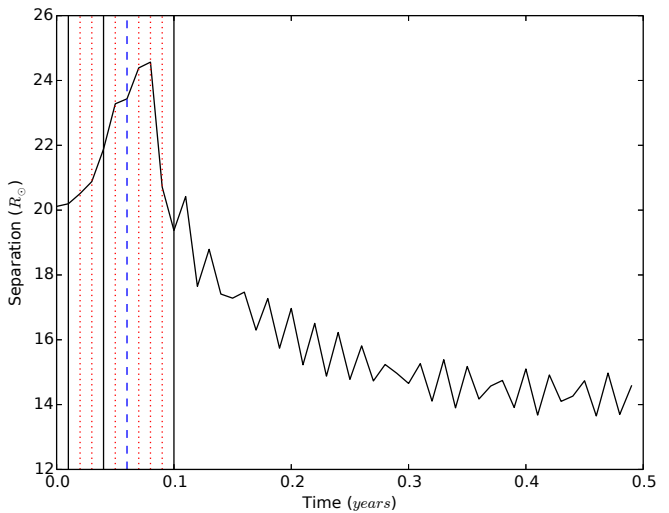}
\caption{The separation of core and companion as a function of simulation time for the {\it Cool} interaction. The vertical lines are time reference points and are described in Section~\ref{ssec:thediscsmassandangularmomentum}.}
\label{separation_timeseries}
\end{figure}
\begin{figure}
\centering
\includegraphics[width=0.5\textwidth]{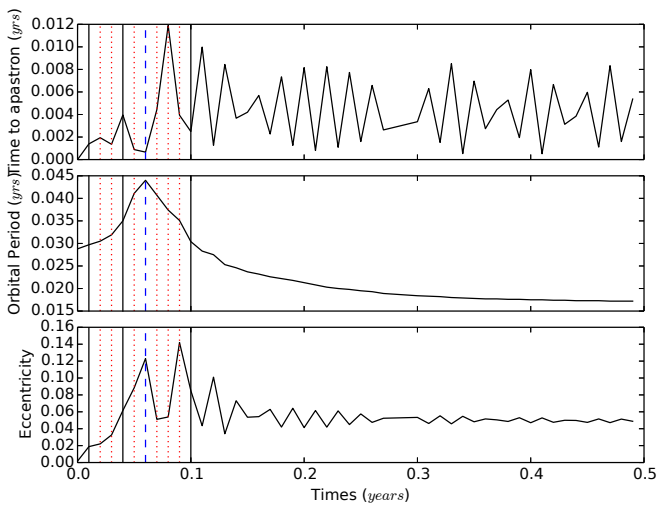}
\caption{Top panel: the time to orbital apastron at each data output is shown in the top panel for the {\it Cool} simulation. Middle panel: the orbital period of the binary. Bottom panel: the eccentricity of the system. The vertical lines are time reference points and are described in Section~\ref{ssec:thediscsmassandangularmomentum}.}
\label{cool_eccentricity}
\end{figure}

\subsection{Determination of the amount of unbound gas}
\label{ssec:determinationoftheamountofunboundmaterial}
\begin{figure}
\centering
\includegraphics[width=0.5\textwidth]{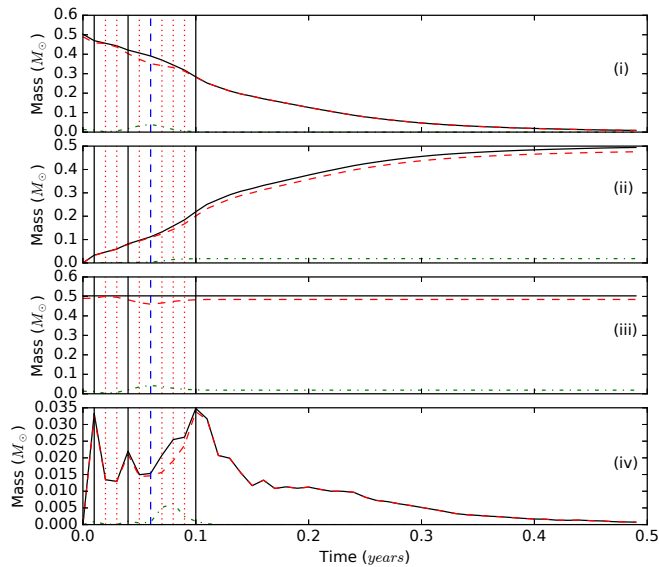}
\caption{(i) Total mass in the computational domain; (ii) cumulative mass leaving the domain; (iii) total mass inside and outside the computational domain; (iv) mass lost  from the computational domain per data output for the {\it Cool} simulation. In all panels, the black line shows total mass, while the dashed red and dash-dotted green lines indicate the bound and unbound mass, respectively. The unbinding event (blue dashed vertical line) is defined as the peak unbound mass within the domain in panels (i) and (iii), while the three major mass loss events, seen as three peaks in panel (iv) are marked by black vertical lines. }
\label{cool_mass_timeseries}
\end{figure}

We calculate the amount of unbound gas by summing the potential, kinetic and thermal energies. This method gives an upper limit to the amount of unbound gas because the thermal energy may be converted into kinetic energy due to expansion, but it may also be lost radiatively. The bound and unbound gas in the grid is shown in Figure \ref{cool_mass_timeseries}. In each panel the red and green lines indicate the bound and unbound mass, respectively, while the black line indicates the sum of the two. In Figure \ref{cool_mass_timeseries} (i) we see a peak in unbound mass in the grid at $0.06$~years due to the core and companion imparting energy to the gas. We call this peak the ``unbinding event" and mark it by a blue vertical dashed line in the figures.  In Figure \ref{cool_mass_timeseries} (ii) we show the amount of bound and unbound gas leaving the computational domain. This was estimated by multiplying the fraction of unbound mass within the boundary layer by the mass lost between code outputs. The boundary layer is defined as the 6, one-cell thick faces at the edge of the computational domain. We implicitly assumed that the unbound mass remained constant between outputs and that the fraction of unbound gas in the boundary layer was representative of the gas within a few cells of the boundary. From these plots we see that the unbound mass from the unbinding event appears to leave the grid $0.01$~to~$0.02$~years later. Summing together the unbound gas fraction within the box and the unbound mass that has left the box (see Figure \ref{cool_mass_timeseries} (iii)) we see that the unbound gas fraction levels off at around $4$ per cent of the total mass of the gas. This translates into $5$ per cent of the fall-back disc mass becoming unbound. The dips in the total unbound mass in the system as seen in Figure \ref{cool_mass_timeseries} (iii) signify that unbound mass becomes bound. 
This is due to gas becoming unbound by interacting with the particles, but later losing energy before leaving the grid either by slowing down because of colliding with bound gas or by adiabatic cooling (we remind that we include the gas' thermal energy in  calculating whether mass is unbound). The mass loss per data output is shown in Figure \ref{cool_mass_timeseries} (iv).

\subsection{Energy and angular momentum conservation}
\label{ssec:energyandangularmomentumconservation}

\begin{figure}
\centering
\includegraphics[width=0.5\textwidth]{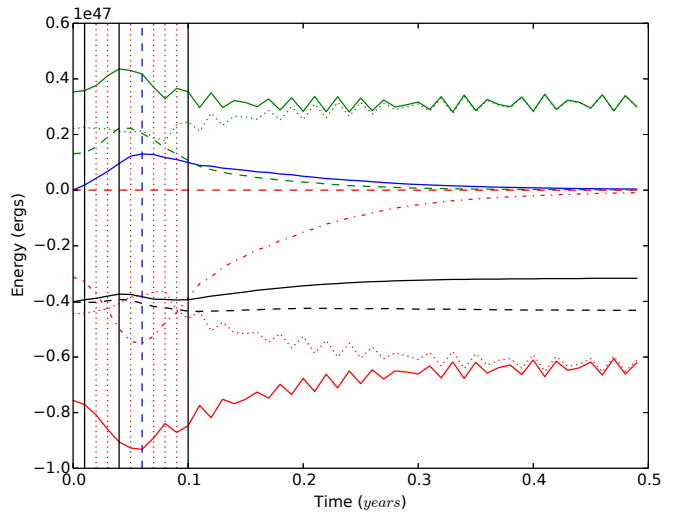}
\caption{Total energies within the box for the {\it Cool} simulation. Plotted are the total energy (solid black line), total energy corrected for mass loss from the box (dashed black line), thermal energy (blue line), kinetic energy of the particles (dotted green line), kinetic energy of the gas (dashed green line), the total kinetic energy (solid green line), the potential energy of the particles with respect to each other (dotted red line), the potential energy of gas on itself (dashed red line), the potential of the particles with respect to the gas (dash-dot red line) and total potential energy (solid red line). The vertical lines are time reference points and are described in Section~\ref{ssec:thediscsmassandangularmomentum}.}
\label{energy_timeseries}
\end{figure}

Since mass flows out of the grid, energy and angular momentum are not conserved. However we have devised an approximate method to check whether the code is conserving energy. The increase in total energy by $8.5 \times 10^{45}$erg or $21$ per cent of the initial total energy of $-4 \times 10^{47}$ (Figure \ref{energy_timeseries}) is due to the loss of bound gas from the grid (which negative energy). We estimated the energy lost from the computational domain through the two computational domain faces parallel ($E_{\rm loss,para}$) and the four faces perpendicular ($E_{\rm loss,perp}$) to the orbital plane, by calculating the average specific total energy within those faces. We then found the fraction of mass in the boundary layer within the faces parallel, $f_{\rm para}$ , and perpendicular, $f_{\rm perp}$, to the orbital plane ($f_{\rm para} + f_{\rm perp} = 1$). Assuming these mass and energy fractions remained relatively constant over the time between data outputs, the energy lost through faces parallel and perpendicular to the orbital plane was estimated in the following way:

\begin{equation}
E_{\rm loss,para} = E_{\rm spec,para} \times f_{\rm para} \times M_{\rm loss}
\label{mperp}
\end{equation}
\begin{equation}
E_{\rm loss,perp} = E_{\rm spec,perp} \times f_{\rm perp} \times M_{\rm loss},
\label{mplane}
\end{equation}

\noindent where $M_{\rm loss}$ is the mass lost between data outputs. This gives the energy loss at the time of each data output to be that shown in Figure \ref{energy_loss}, upper panel. From this we see that most of the energy leaves the system perpendiculat to the orbital plane. Adding the energy lost calculated with this method to the energy within the box, brings the total energy to a fairly constant value (the black dashed line in Figure \ref{energy_timeseries}), with a maximum excursion of ${\sim} 6$ per cent from the initial total energy.

\begin{figure}
\centering
\includegraphics[width=0.5\textwidth]{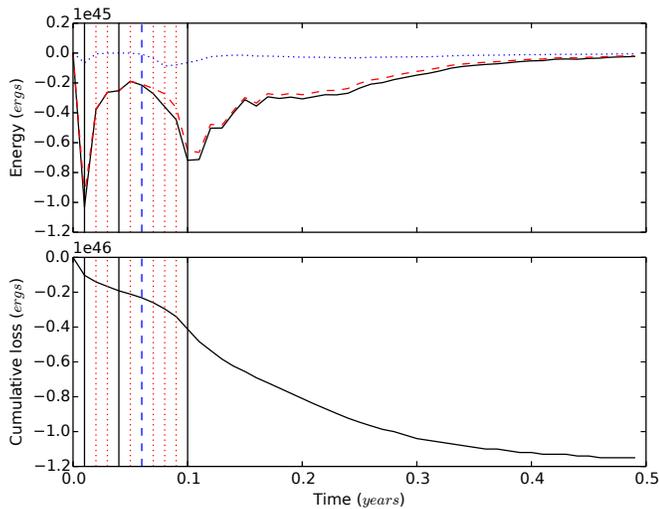}
\caption{Energy loss per data output for the {\it Cool} simulation (top panel) as estimated using Eqs.\eqref{mperp} and \eqref{mplane}. The vertical lines are time reference points and are described in Section~\ref{ssec:thediscsmassandangularmomentum}. The blue dotted line indicates energy loss through the four computational domain faces perpendicular to the orbital plane and the red dashed line indicates energy loss through the two faces parallel to the orbital plane. The black solid line indicates the total energy lost per data output. In the bottom panel the cumulative energy loss from the computational domain is shown.}
\label{energy_loss}
\end{figure}

Next we considered the conservation of angular momentum ($\boldsymbol{\mathrm{L}}$), which was calculated using:

\begin{equation}
\boldsymbol{\mathrm{L}} = \boldsymbol{\mathrm{r}} \times \boldsymbol{\mathrm{p}}
\label{angular_momentum}
\end{equation}

\noindent where $\boldsymbol{\mathrm{r}}$ is the vector location of the centre of each cell with respect to the  centre of mass of the system and $\boldsymbol{\mathrm{p}}$ is the linear momentum of the gas contained in that cell. Figure \ref{angular_momentum_timeseries} shows that the $z$-component of the total angular momentum (which is by far the dominating component due to the geometry of the system) declines as gas leaves the computational domain. We accounted for the lost angular momentum in the same manner as we did for the energy. The dotted black line in Figure \ref{angular_momentum_timeseries} shows that the angular momentum, once accounted for losses from the grid, is conserved at the 8 per cent level, where the error bar on this conservation estimate is relatively large due to the approximate calculation of the energy and angular momentum content of the departed gas. The angular momentum of the particles suffers a temporary increase corresponding to the orbital separation increase experienced early in the simulation (Figure \ref{separation_timeseries}). Overall the particles lose approximately 20 per cent of their angular momentum.

\subsection{Impact of gas temperature on the particles' in-spiral}
\label{ssec:impactofthegastemperatureontheparticlesinspiral}
\begin{figure*}
\centering
\includegraphics[width=0.9\textwidth]{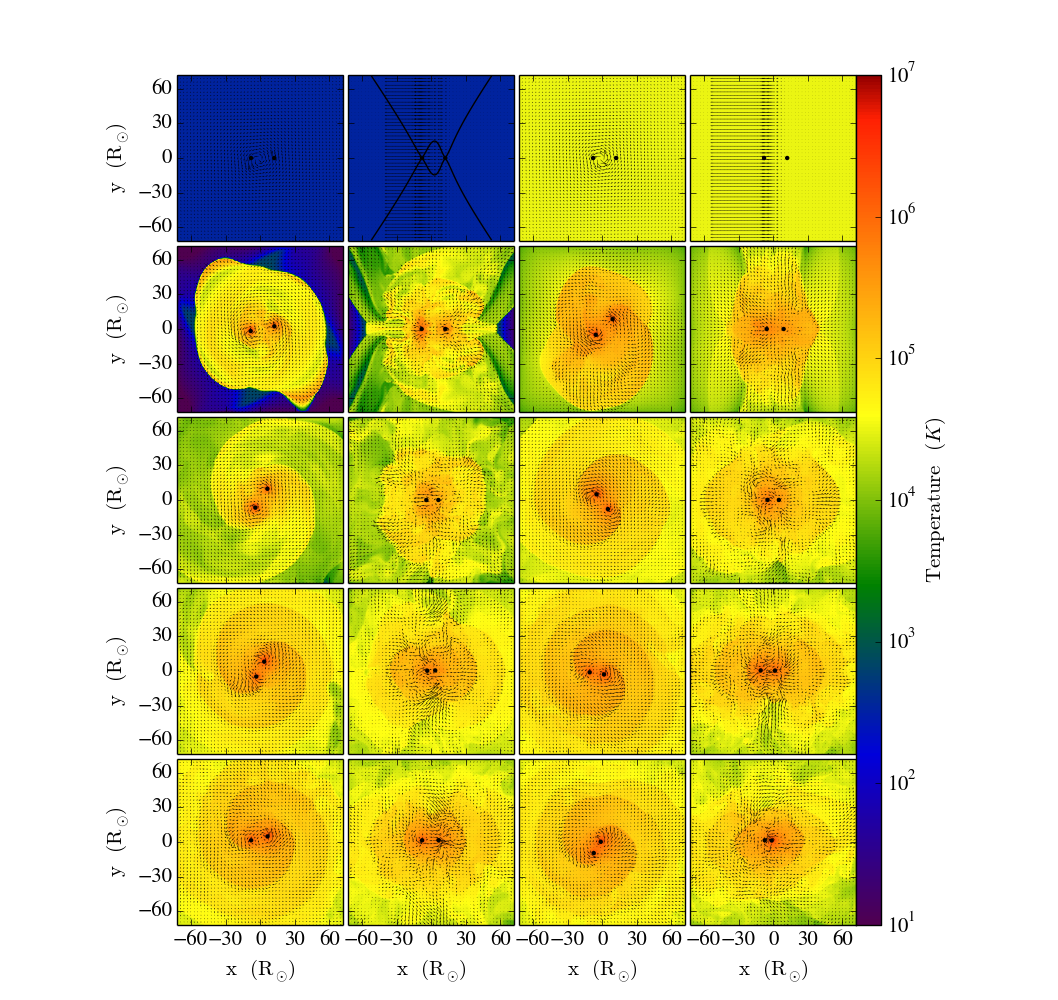}
\caption{The temperature evolution for a slice along the orbital plane (first and third columns) and perpendicular to the orbital plane (second and last columns), at the beginning of the simulation (top row), at $0.01$~years (second row), at $0.02$~years (third row), $0.03$~years (last row). First two columns: the {\it Cool} simulation; Last two columns: the {\it Hot\_tuned} simulation.}
\label{temperature_evolution}
\end{figure*}
As we have explained in Section~\ref{ssec:thediscsmassandangularmomentum} we do not know what the temperature of the returning gas is, but it likely would play a role on the dynamics of the returning envelope.  We also wonder what effect the temperature of the gas in the immediate vicinity of the in-spiralling particles has on the gravitational drag force. The drag force is related to the density and velocity contrast of the gas bathing the particles, but also on the Mach number of the particles which is a function of temperature \citep{ostriker_dynamical_1999}.

Our {\it Cool} simulation started with a gas temperature of $350$~K. Using this initial temperature at $t=0$, the initial orbital velocities of the particles are highly supersonic.
We calculated the Mach numbers of the particles using the average sound speed of the gas within a $1$~R$_\odot$ sphere around the particles. However, suspecting that some gas becomes trapped in the potential well of the particles, effectively travelling with them thus lowering the relative velocities, we also calculated the sound speed of the gas using a box of $3^3$ cells, centred on the cell located two cells in front of the one containing the particle. In practice these two methods returned similar answers. The particles' velocities were corrected to be relative to the average velocity of the gas used to calculate the sound speed. The initial Mach numbers of the $0.39$~M$_\odot$ core and of the $0.6$~M$_\odot$ companion were $\mathcal{M} = 79$ and $\mathcal{M} = 43$, respectively, dropping to below  unity at 0.01~yr of the simulation and maintaining values of $\sim$0.2 thereafter.  The strong shock heating quickly results in a temperature profile with values between $\sim10^4$ and $\sim 10^6$~K, only 0.01-0.02~yr after the start of the simulation, when the particles are just starting to interact with the increased density of the newly formed disk (Fig.~\ref{temperature_evolution}).

We ran two additional simulations with a higher initial temperature of 35\,000~K and the same initial density of $10^{-6}$~g~cm$^{-3}$. 
With this temperature the Mach numbers of the core and companion at the beginning of the hot simulations were lower: 5 and 10, respectively, dropping to 0.5 at time 0.01~yr and remaining around this value thereafter. 
The behaviour of the Mach number in the two simulations is therefore similar, transitioning to subsonic before 0.01~yr of simulation, releasing the suspicion that this may influence the in-spiral \citep{ostriker_dynamical_1999} and leaving other factors such as density and velocity contrast to be investigated. 

For the first of the two ``hot" simulations with an initial temperature of 35\,000~K, which we nickname \emph{Hot\_fast}, we retained the same initial velocity setup (see Section 2.). However, maintaining the same velocity distribution with higher temperature results in less fall-back mass ($0.28$~$M_\odot$ vs. $0.38$~$M_\odot$ for the {\it Cool} simulation). Therefore, we also ran a second high temperature simulation, but with a slower velocity profile, where gas velocities on the particles' orbital plane were given $0.4 \times v_{\rm Kep}$ instead of a value of $0.75 \times v_{\rm Kep}$. We nicknamed this simulation \emph{Hot\_tuned}. In this way we tuned the simulation to generate a fall back disc mass ($0.34$~$M_\odot$) that was closer to that in the \emph{Cool} simulation ($0.38$~$M_\odot$).

The unbinding efficiencies of the simulations are measured as described in Section 3.2. The amount of unbound gas in \emph{Hot\_fast} and \emph{Hot\_tuned} is 5 per cent and 4 per cent, respectively, virtually the same as for the \emph{Cool} simulation. 

\begin{figure}
\centering
\includegraphics[width=0.5\textwidth]{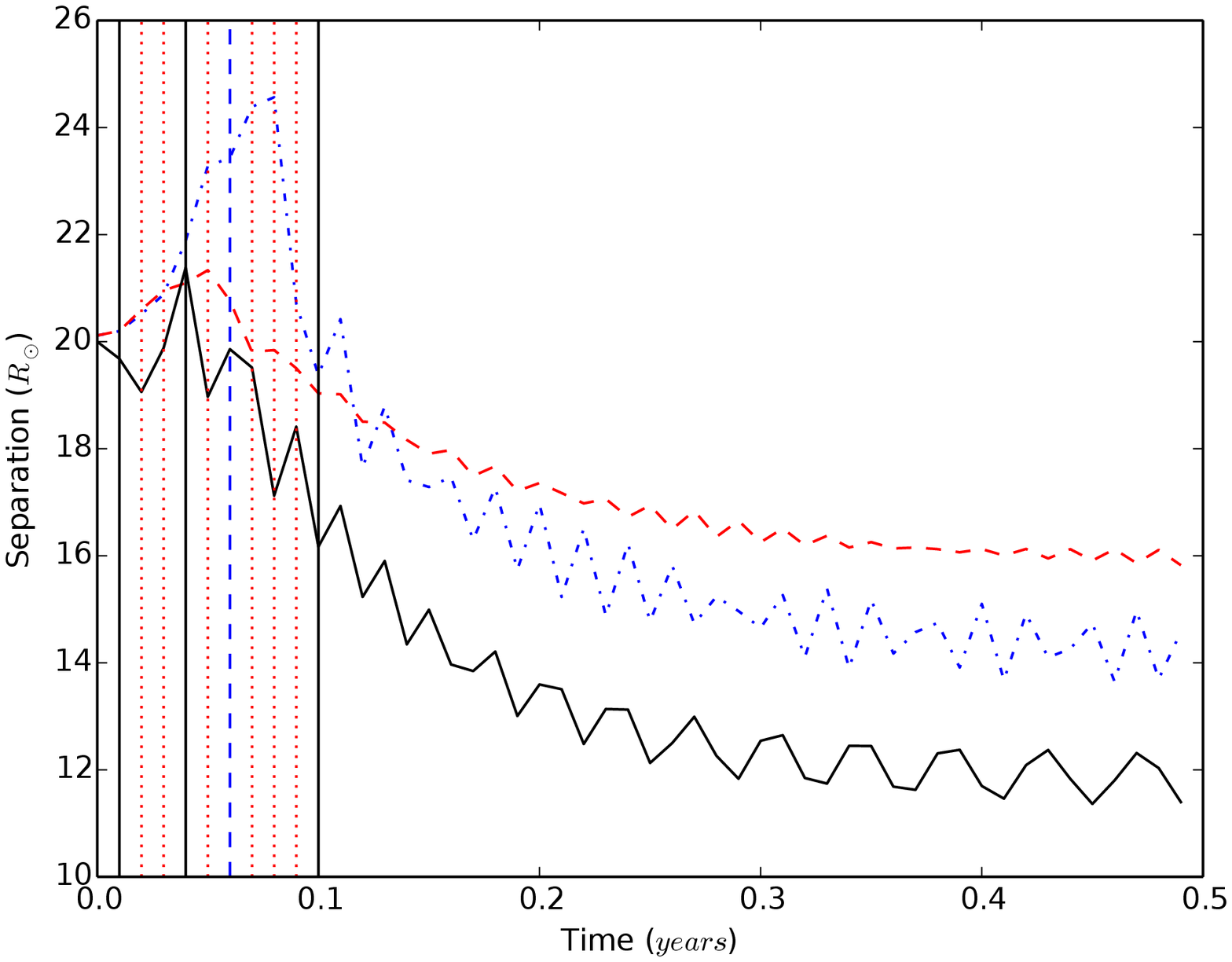}
\caption{Evolution of the orbital separation for simulations with different initial conditions as described in the text. The dot-dashed blue line is for the \emph{Cool} simulation (this is the same as in Fig.~\ref{separation_timeseries}), the dashed red line is for the \emph{Hot\_fast} simulation and the solid black line is for the \emph{Hot\_tuned} simulation.}
\label{separation_comparison}
\end{figure}

We next compare the separation evolution of our three simulations (Figure \ref{separation_comparison}). In the \emph{Cool} simulation the particles in-spiral more than in the \emph{Hot\_fast} simulation, but less than the \emph{Hot\_tuned} simulation. We ascribe the difference in in-spiralling behaviour between the \emph{Cool} and \emph{Hot\_fast} simulation to the different disc masses (0.38 and 0.28~\msun, respectively) rather than to the different initial temperatures, where the lighter disc promotes less in-spiral. This conclusion is drawn based on the fact that looking at simulations \emph{Cool} and \emph{Hot\_tuned}, which have closer disc masse values (0.38 and 0.34~\msun), but different initial temperatures (350 and 35\,000~K) the total amount of in-spiral is similar. The only difference in the in-spiral history of these two simulations is the initial orbital expansion. The \emph{Hot\_tuned} simulation results in less initial expansion due to the fact that the initially slower gas does not accelerate the particles. We conclude that the initial temperature is not a major factor in deciding the rate of in-spiral in the fall-back event for the cases studied. This is likely due to the fact that the temperature profiles converge to similar values soon after the start of the simulation (Figure~\ref{temperature_evolution}).

\citet{Kuruwita2015} tested a range of additional temperatures, up to 175\,000~K. With such a high temperature the pressure is such that much of the material is evacuated from the simulation domain early in the simulation and the mass of the fall-back disks is much reduced (0.2~\msun), necessitating an even greater alteration of the velocity profile in order to achieve the same disk mass as the other simulations. Changing the isothermal temperature value at $t=0$, while adjusting the velocity profile to obtain the right fall back disk mass does not have a direct consequence on the in-spiral pattern and lack of substantial unbinding. However, these tests act as a reminder that the temperature profile would play a role in a real in-fall because it has a direct effect on the pressure.

\subsection{Numerical resolution}
\label{ssec:numericalresolution}

A resolution of 256$^3$ was deemed by the convergence tests of Passy et al. (2012) to be sufficient for the common envelope they simulated. In their simulation one cell corresponded to 1.7~\rsun. With a smoothing length of 1.5 cells, their final separation of $\sim$20~\rsun, was 8 times the linear resolution multiplied by the smoothing length. Our cell size is three times smaller than in Passy et al. (2012), or 0.57~\rsun\ and the final separation of the \emph{Cool} simulation is therefore 16 times the linear resolution multiplied by the smoothng length of 1.5 cells. This releases the suspicion that the resolution limits the in-spiral. 

We also carried out an additional resolution test by repeating the \emph{Hot\_fast} simulation with a higher resolution of 384$^3$. The behaviour of the orbital separation is almost identical to that observed in Fig.~\ref{separation_comparison} (red dashed line) for the lower resolution case: we observe a slightly lesser initial out-spiral of the orbit for the higher resolution simulation, reaching 20.8~\rsun\ instead 21.3~\rsun\ and a slightly smaller final separation of 15.8~\rsun\ instead of 16.0~\rsun. The unbinding efficiency is not significantly affected by the resolution, with the high resolution simulations unbinding $5$\% of the fall back material. { This test (though not a proper convergence test) gives some assurance that resolution does not greatly affect the outputs of our simulation. We discuss this topic further in Section~\ref{ssec:furtherconsiderations}.}

\section{Discussion}
\label{sec:discussion}


Although these simulations are only marginally better than toy models, they do inform our intuition on the role a fall-back disk may have in the context of the common envelope simulation.
All three simulations reduce effectively the orbital separation. The simulations with the smallest disc mass reduce the separation the least. All simulations are instead inefficient in unbinding further envelope gas, with efficiencies at the 5 per cent level, independent of initial setup. These results echo what is observed in the common envelope simulations of Passy et al. (2012), where the extreme orbital separation decrease witnessed in the in-spiral, results in only $\sim10$ per cent of the envelope being unbound.

Taking these results at face value we would conclude that further fall back events would have to take place until either the envelope is unbound or the core and companion merge. Below we therefore calculate how many fall-back events are needed to achieve the ejection of the envelope and compare this number with how many fall back events would result in an orbital separation commensurate with observations. In Section~\ref{ssec:furtherconsiderations} we place our fall-back model in the context of additional phenomena which likely take place at the end of the common envelope rapid in-spiral.

\subsection{The time to the next fall-back event and the number of fall-back events}
\label{ssec:timetothenextfallbackevent}

We calculate the ballistic time it would take gas leaving the computational domain after interacting with the particles to return into the computational domain for a second fall-back interaction. We use the data output at a time of 0.08 years from the beginning of the simulation, when a substantial amount of gas is leaving the domain.  We assume that the velocity of the gas in each cell within the boundary of the grid is directed radially outward. Although all gas within the boundary has a velocity component directed outward, using the total velocity modulus will result in an upper limit of the return times. As we shall see this is not important because the times calculated are short.  Using this velocity and only the acceleration due to the gravity of the core and companion, we integrate the change in radial distance over time steps of one week and determine the time when the gas parcel comes back to the same position. 

We have plotted the estimated fall-back times for each gas parcel vs. the mass of the parcel in Figure \ref{fallback_time} (for simulation {\it Cool}).  
\begin{figure}
\centering
\includegraphics[width=0.5\textwidth]{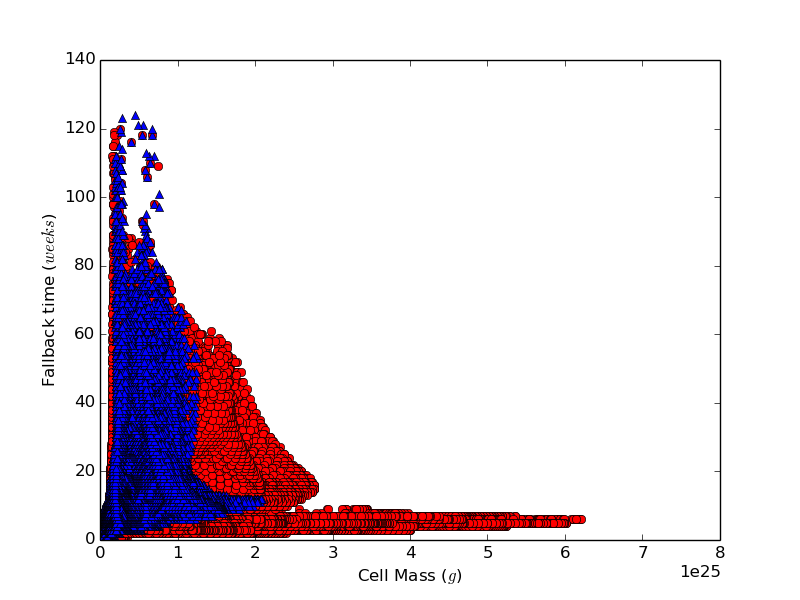}
\caption{Estimated upper limits for the fall-back time of bound gas parcels leaving the grid boundary at 0.08 years in the simulation, plotted against the mass of the cell. Gas parcels leaving the domain above and below the orbital plane are shown by blue triangles, while parcels leaving the domain through faces perpendicular to the orbital plane are shown by red circles.}
\label{fallback_time}
\end{figure}
Here we can see that the upper limits to the return time to the computational domain has a large spread of values, from a few weeks to approximately two and a half years. The bulk of the mass loss happens through faces perpendicular to the orbital plane (red circles in Figure \ref{fallback_time}), as expected, with the bulk of the return times having upper limits between a few and 80 weeks. In conclusion the next fall-back event is likely to happen very rapidly after the first.

The common envelope SPH simulations of \citet[][with an identical setup to the simulation of Passy et al. (2012) discussed here, but extending the simulation time to $10$~yrs]{Reichardt2016} also demonstrate that $\sim$3 years after the beginning of their simulation some bound envelope mass is returning to the centre. This return will however be slowed down by building pressure. The dynamical return we envisage, may therefore take substantially longer (see further discussion on this point in Sec.~\ref{ssec:furtherconsiderations}). 

With knowledge of the efficiencies with which a fall-back event reduces the orbital separation and unbinds further mass, we can calculate the number of required events to reduce the separation to be within the observed values as well as the number of events required to unbind the envelope. 

To calculate the number of fall back events required to reduce the separation, we use a target separation value of $4$~R$_\odot$, based on the work of \citet{zorotovic_post-common-envelope_2010} and \citet{de_marco__2011}.
We know that each consecutive fall back event will have a less massive disk than the one preceding it and will become less efficient in reducing the orbital separation. Below we use a constant efficiency thereby calculating a lower limit for the number of fall back events necessary to bring our system to the observed separations. With this assumption the orbital separation after $n$ fall-back events is given by:

\begin{equation}
a_n = a_0(1-\epsilon)^{n},
\label{orbital_separation}
\end{equation}

\noindent where $a_0$ is the initial orbital separation, $20$~R$_\odot$, $\epsilon$ is the in-spiralling efficiency, or the reduction in orbital separation divided by the separation at the beginning of each event, $n$ is the number of fall-back events we want to know and $a_n$ is the orbital separation after $n$ fall-back events, or 4~\rsun.  The efficiency $\epsilon= 0.43$ based on our simulations shown in Figure \ref{separation_comparison} (where we have used here the in-spiral from the maximum separation of the \emph{Cool} simulation). Therefore the minimum number of necessary fall back events to bring the system to within observed separations is 3, but it would be larger if the in-spiralling efficiency decreased at each event. 

Applying this same reasoning to calculate the number of necessary fall back events to unbind the entire envelope we use the above equation, but with envelope mass instead of separation and with the unbinding efficiency of 0.05, as calculated in Sec.~\ref{ssec:determinationoftheamountofunboundmaterial}.
For the calculation we use an initial mass of 0.44~\msun\ and a final mass of 0.1~\msun. The latter value is based on the assumption that a certain amount of mass can remain in orbit around the binary. This is highly likely to be the case. \citet{tocknell_constraints_2014} calculated that there is a spread in the specific angular momenta of the bound gas in the common envelope simulation of Passy et al. (2012), with some of the gas potentially coming to rest at an orbital separation larger than the orbital separation of the cores. Although an estimate of this mass will have to wait for a better calculation, we note that \citet{Kashi2011} used 0.2~\msun\ in their somewhat {\it ad hoc} consideration of fall back disks. With our final mass value and a constant unbinding efficiency of 0.05 we estimate that 29 events will be necessary to unbind the envelope. If the efficiency were reduced to 0.01 (due to not considering thermal energy in the amount of unbound mass), then the number of fall back events would become 147. These two numbers would grow to 73 and 376, respectively if the left over disk mass was 0.01~\msun\ instead of 0.1~\msun.  All these numbers are upper limits because as the separation decreases, equal $\Delta r$ changes deliver increasing orbital energy.  

Based on the estimates and considerations above, { and even considering the approximate nature of our model, it seems} unlikely that the number of fall back events would lead to  the correct separation {\it and} the unbinding of sufficient envelope. More likely the two cores would merge before sufficient envelope could be unbound. 

\section{Final considerations and conclusion}
\label{ssec:furtherconsiderations}

We have carried out a set of simulations under the assumption that a torus of gas falls back onto the post in-spiral binary. The binary mass and separation, as well as the disk's mass and angular momentum were tuned to match the central binary, the bound mass and angular momentum, respectively, measured at the end of a common envelope simulation by Passy et al. (2012). Our simulations show that a fall-back disk such as the one we envisage would not  lead to unbinding sufficient mass. We therefore conclude that, alone, such a process could not achieve substantial further unbinding {\it and} orbital reduction.

One may wonder whether dynamically-returning envelope gas will in fact form such a disk. Ivanova et al. (2013) reviewed the idea that at the end of the dynamical in-spiral, a ``self-regulated" phase allows the final separation to be reduced further over a much longer thermal time scale. During this phase the energy deposited by the binary is radiated away by whatever envelope is still bound to the giant core, the giant contracts, the density surrounding the cores  increases and the binary in-spirals further. 

Before the thermally-regulated phase, but after the fast-in-spiral modelled by 3D simulations, there has to be a phase during which outflowing, bound gas returns towards the binary dynamically and settles into a temporary equilibrium configuration, which will then evolve over thermal timescales. The amount of bound envelope will dictate the conditions of this dynamical return phase.
If a large fraction of the envelope is still bound, it will return dynamically, but it may be halted by the pressure that builds while the large mass of envelope effectively is re-forming a star. It is possible that under these conditions a disk or torus may not form.


We suggest, however, that the doughnut geometry of the expanding common envelope observed, at least on a large scale, in Passy et al. (2012; see also, e.g., Sandquist et al. 1998) and the high angular momentum of the returning gas would result in a toroidal structure. In addition, the rotation profile of the gas that interacts with the binary during the fall-back sensitively dictates how much orbital energy is transferred to the in-falling gas \citep{ostriker_dynamical_1999}, and how much more in-spiral takes place during the end of the dynamical phase. 
Without a full simulation it is difficult to determine how the end of the dynamical phase would look like, or how much farther the binary would in-spiral before the system settled into a slower, thermal-timescale phase.

The addition of recombination energy in simulations has been suggested to be an effective way to unbind more envelope gas, at least under some circumstances \citep[][]{ivanova_role_2015,nandez_recombination_2015}. 
If recombination energy increases gas unbinding, the mass in a presumed fall-back disk or torus would be lower and an interaction such as the one envisaged here would result in less in-spiral. It is often found that the truth is in the middle. It is possible that the correct combination of unbinding and in-spiral is a cocktail of recombination energy (which boosts unbinding), some dynamic fall-back interaction (which promotes further in-spiral), a thermal readjustment phase and tidal action by a low-mass, circumbinary, left-over disk as envisaged by \citet{Kashi2011}.

We do not include magnetic fields in our simulations. \citet{Regos1995}  derived analytically that at the end of the common envelope in-spiral, field strengths of a few hundred Gauss would develop via the winding action of the binary \citep[see also][]{tocknell_constraints_2014}. As the envelope expands, the field strength would possibly decrease, but this could increase again if most of the envelope fell back. Such strong field could give some extra buoyancy to the returning gas, but would also provide a viscous force \citep{Wardle2007} allowing a more efficient redistribution of the angular momentum. 

Viscosity is what redistributes angular momentum. Convective motion promoted by the in-spiral (Ohlmann et al. 2016) alongside intensified magnetic fields should increase the viscosity in the common envelope. Viscosity in our simulations has a numerical origin, and we gauged it to be very low using the criterion of \citet{Federrath2011}, who showed that more than 30 cells per Jeans length are needed to lower the numerical viscosity (we have ~100 cells in the key regions near the particles). Viscous forces are therefore poorly reproduced in hydrodynamic simulations (see the discussion by Rasio and Livio (1996); their section 4.2) something that casts doubt on simulation results, particularly pertaining end of the dynamical phase, when in-falling gas interacts with itself.

Despite this shortcoming, simulations of the common envelope phase are fast improving. In the last two years more codes, encompassing additional physics, have been applied to the problem \citep[][]{nandez_recombination_2015,Hwang2015,Ohlmann2016,Reichardt2016}. Simulating a common envelope for many more dynamical times following the post-fast-in-spiral phase is not beyond the realm of possibility, something that will contribute to answer the question of the actual geometry of the returning gas and its effect on the outcome of the common envelope interaction.

\section*{Acknowledgments}

We thank Paul Ricker and Daniel Price for useful comments to the Master thesis of Rajika Kuruwita, which constitutes the basis of this article. { We are thankful to Jean-Claude Passy for access to his simulations that provided the basis for our work and Daniel Macdonald for help with the orbital parameters calculation. We are finally thankful to an anonymous referee for a series of comments, which provided much of the basis for the discussion in Section~\ref{ssec:furtherconsiderations}}. This research was undertaken with the assistance of resources from the National Computational Infrastructure (NCI), which is supported by the Australian Government, as well as the swinSTAR supercomputer at Swinburne University of Technology. Also the Australian Government and the financial support provided by the Macquarie University Master of Research postgraduate research program. This  work was  supported in  part by an Australian Research Council Future Fellowship to De~Marco (FT120100452) and an Australian Research Council Discovery Project (DP12013337).

\bibliographystyle{mn2e}
\bibliography{bibliography}

\end{document}